\documentclass[aps,reprint,superscriptaddress, nobalancelastpage, raggedbottom]{revtex4-2} 
 \usepackage{ragged2e}
\usepackage{amsmath}
\usepackage{amssymb}
\usepackage{physics}
\usepackage{booktabs}
\usepackage[normalem]{ulem}
\usepackage{graphicx}%
 \usepackage{setspace}
 \usepackage[skip=5pt]{caption}
\usepackage{dcolumn}
\usepackage{bm}
\usepackage[dvipsnames]{xcolor}

\begin{document}

\title{Designing Quaternary Hydrides with Potential High T$_c$ Superconductivity}

\author{Adam Denchfield}
\affiliation{Department of Physics, University of Illinois Chicago, Chicago, Illinois 60607, USA}
\author{Russell J. Hemley}
\affiliation{Department of Physics, University of Illinois Chicago, Chicago, Illinois 60607, USA}
\affiliation{Department of Chemistry, University of Illinois Chicago, Chicago, Illinois 60607, USA}
\affiliation{Department of Earth and Environmental Sciences, University of Illinois Chicago, Chicago, Illinois 60607, USA} 
\author{Hyowon Park}
\affiliation{Department of Physics, University of Illinois Chicago, Chicago, Illinois 60607, USA}
\affiliation{Materials Science Division, Argonne National Laboratory, Lemont, Illinois 60439, USA}


%
\date{July 1st, 2024}

\begin{abstract}
  While hydrogen-rich materials have been demonstrated to exhibit high T$_c$ superconductivity at high pressures, there is an ongoing search for ternary and quaternary hydrides that achieve such high critical temperatures at much lower pressures. First-principles searches are impeded by the computational complexity of solving the Eliashberg equations for large, complex crystal structures. Here, we adopt a simplified approach using electronic indicators previously established to be correlated with superconductivity in hydrides. This is used to study complex hydride structures, which are predicted to exhibit promisingly high critical temperatures for superconductivity. In particular, we propose three classes of hydrides inspired by the FCC RH$_3$ structures that exhibit strong hydrogen network connectivity, as defined through the electron localization function. The first class [RH$_{11}$X$_3$Y] is based on a Pm$\overline{3}$m structure showing moderately high T$_c$, where the T$_c$ estimate from electronic properties is compared with direct Eliashberg calculations and found to be surprisingly accurate. The second class of structures [(RH$_{11}$)$_2$X$_6$YZ] improves on this with promisingly high density of states with dominant hydrogen character at the Fermi energy, typically enhancing T$_c$. The third class [(R$^1$H$_{11}$)(R$^2$H$_{11}$)X$_6$YZ] improves the strong hydrogen network connectivity by introducing anisotropy in the hydrogen network through a specific doping pattern. These model structures and the design principles provide the enough flexibility to optimize both T$_c$ and the structural stability of complex hydrides.
  
\end{abstract}

\maketitle

The synthesis of materials that superconduct at room temperature and atmospheric pressure has been a long-standing dream since the discovery of superconductivity \cite{onnes1911superconductivity}. The theoretical predictions of high T$_c$ superconductivity in metallic hydrogen \cite{ashcroft1968metallic, ginzburg1972problem} and its subsequent prediction in dense hydrogen-rich materials \cite{ashcroft2004hydrogen} have been realized experimentally \cite{somayazulu2019evidence, drozdov2019superconductivity, snider2021synthesis, kong2021superconductivity, troyan2021anomalous} (see \cite{hilleke2022tuning} for a review), alongside theoretical work. Notable in this regard are predictions of extremely high-T$_c$ superconductivity in LaH$_{10}$ \cite{liu2017potential, peng2017hydrogen} which were found to agree closely with experiment \cite{somayazulu2019evidence, drozdov2019superconductivity}, followed by later experiments on the YH$_n$ system \cite{kong2021superconductivity, snider2021synthesis, troyan2021anomalous} and ternary hydrides \cite{semenok2021superconductivity, song2023stoichiometric, chen2024high}. Several more recent studies have proposed ternary or more complex hydrides as vehicles to lower the pressure needed for very high-temperature superconductivity \cite{liang2021prediction, sun2022prediction, koshoji2022prediction, dolui2023feasible, shutov2024ternary, zheng2024prediction, liang2024design}.

Many such studies start with the highest T$_c$ stoichiometric compounds and examine alloying effects that lower the pressure necessary to stabilize their structures. An alternative approach is to begin with hydrogen-rich MH$_3$ units that are already stable at ambient conditions and embed them in supercells designed to maximize T$_c$, as alluded to in Ref. \cite{ashcroft2004hydrogen} and explored in a recent computational study of [Al,Ga]H$_3$ alloys \cite{liang2024design}. The FCC rare-earth trihydride (RH$_3$; R=Sc,Y,La-Lu) compounds have strong electron-phonon coupling (EPC) \cite{racu2006strong, shao2021superconducting, lu2023electron, ferreira2023search, dangic2023ab} associated primarily with octahedrally coordinated hydrogen, as in palladium hydrides \cite{meninno2023ab}. Notably, experiments on YH$_3$ indicate that the FCC phase can be metastabilized with doping \cite{van2001insulating, kataoka2018stabilization}, ball milling \cite{kataoka2021face}, and quenching \cite{kataoka2022origin}. In FCC RH$_3$, doping the rare earth \cite{villa2022superconductivity} and substituting nitrogen for octahedral hydrogen in LuH$_3$ \cite{ouyang2023superconductivity, hao2023first} have both been predicted to increase this electron-phonon coupling. Notably, evidence of superconductivity in near-ambient conditions has been reported for FCC LuH$_{3-x}$N$_y$ \cite{dasenbrock2023evidence, salke2023evidence} with 0.8\% N by weight. Other experiments indicate that the FCC phase of N-doped LuH$_{3-\delta}$ can be recovered to ambient conditions \cite{li2024stabilization} and displays nearly-flat electronic bands just under the Fermi energy E$_F$\cite{liang2023observation}. Nearly-flat bands were first predicted in Fm$\overline{3}$m Lu$_8$H$_{23}$N \cite{denchfield2024electronic} which is predicted to be dynamically stable at 2 GPa when including nuclear quantum effects \cite{denchfield2024quantum}.

Inspired by the above work, we generalize the idea of doping the metal atom in RH$_3$ to include small amounts of doping of the octahedral hydrogen with Lewis bases [O,N,S,P]. As the space of possible quaternary structures and stoichiometries is huge, we use FCC RH$_3$ and the structural motif from our previous work \cite{denchfield2024electronic} to design model structures with relatively few degrees of freedom. Specifically, we propose three model structures of increasing complexity that are engineered to have dominant hydrogen states at the E$_F$ that emulate metallic hydrogen \cite{ashcroft2004hydrogen}. These structures are summarized in Fig. \ref{fig:threestructs}. When discussing specific examples, we use these structures as starting points and perform DFT variable-cell relaxations. We estimate T$_c$ using a linear correlation observed between Eliashberg T$_c$ and a quantity $\Phi$ based on electronic properties of the hydrogen sublattice in hydrides (see \textit{Materials and Methods}). We also perform frozen phonon calculations near ambient pressures to examine the dynamical stability of selected structures. 

\section*{Results}

\subsection*{Pm$\overline{3}$m RH$_{11}$X$_3$Y Structure}

\begin{figure}
  \centering
   \includegraphics[width=8.7cm]{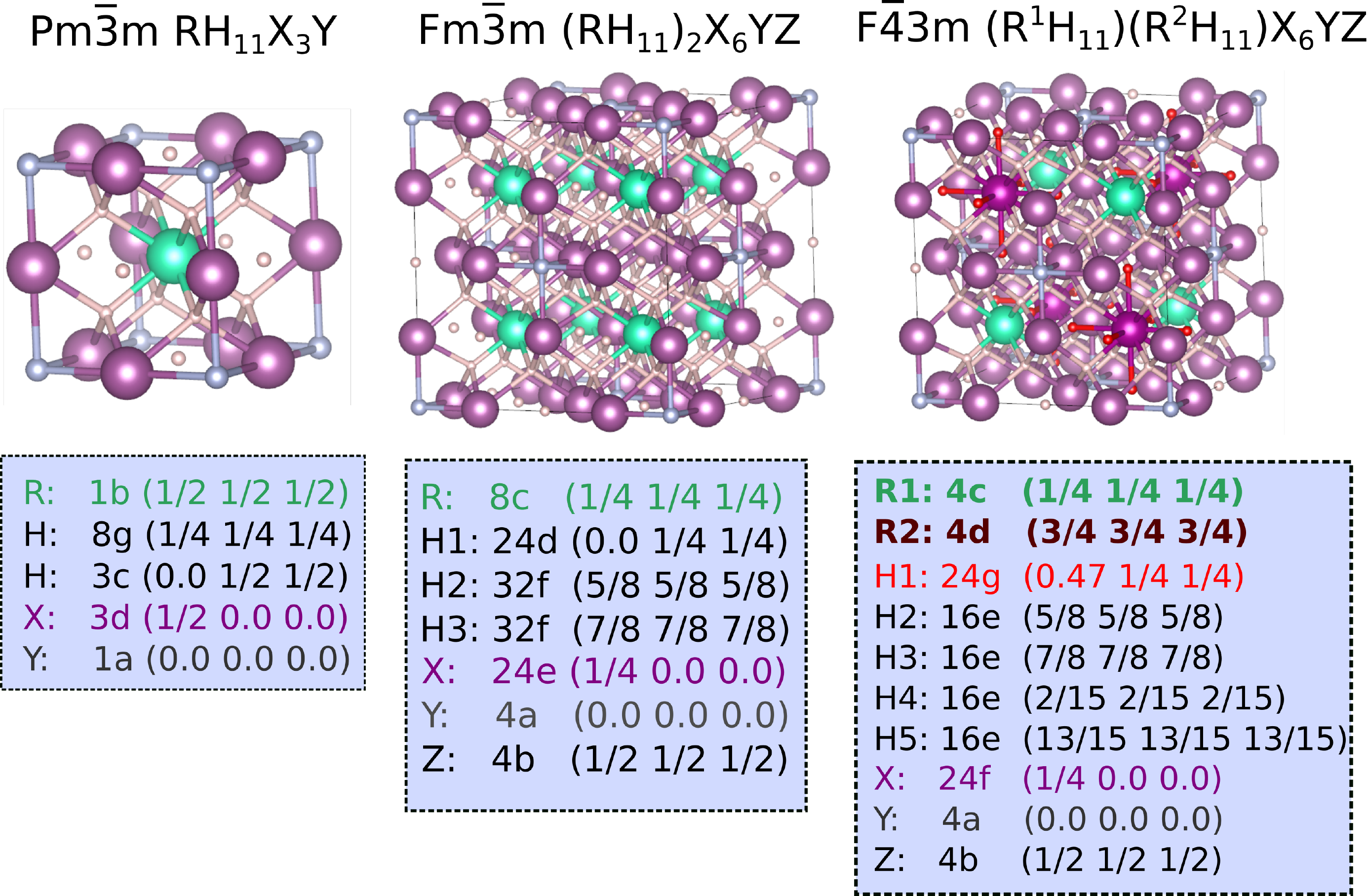}
     \caption{Three structures of increasing complexity that serve as model structures for quaternary (or even more complex) hydrides.}
   \label{fig:threestructs}
 \end{figure} 

\begin{figure}
   \centering
     \includegraphics[width=1.0\linewidth]{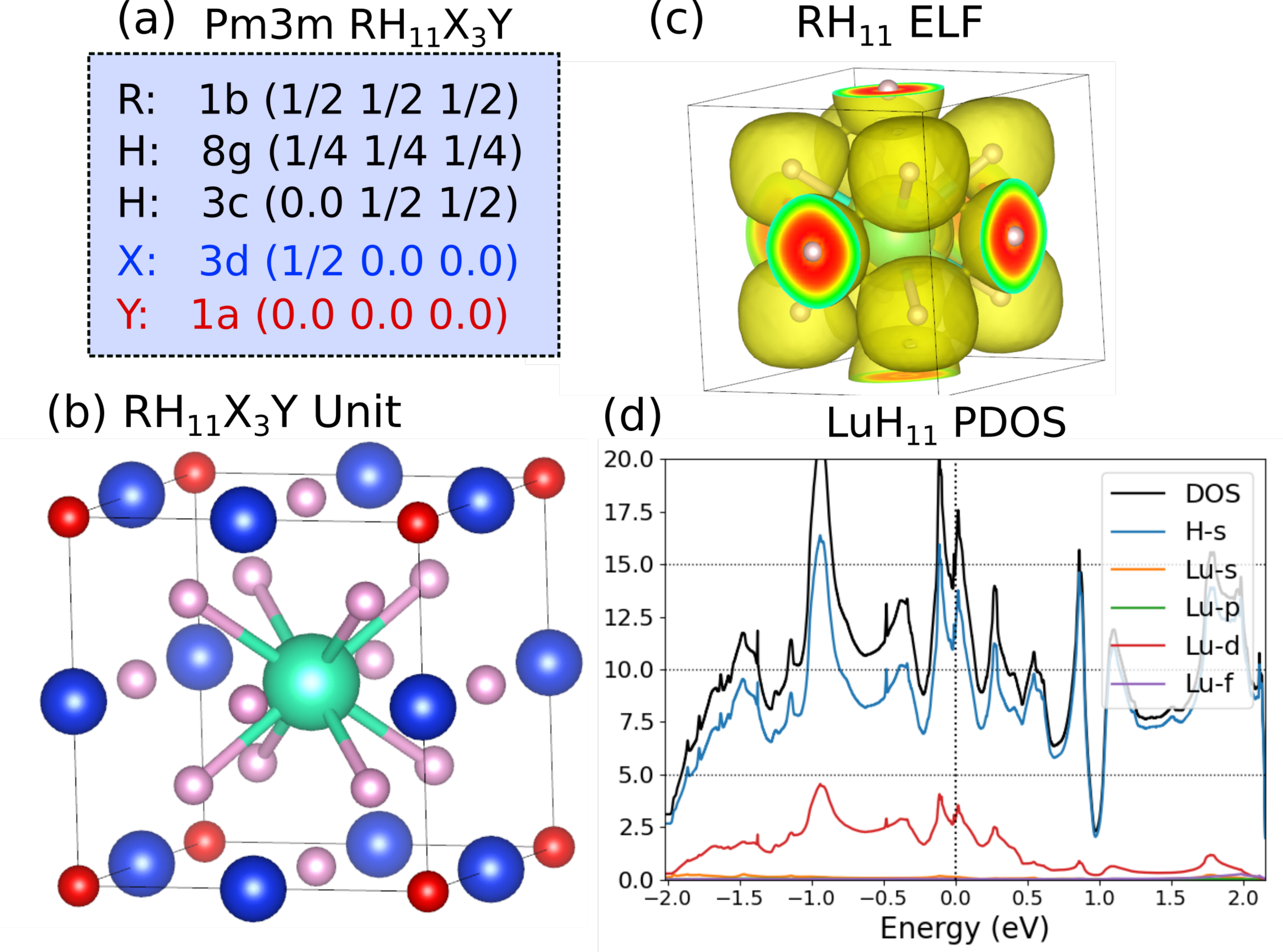}
   \caption{Proposed RH$_{11}$X$_3$Y structure. (a) Structural parameters, (b) Visualization of the RH$_{11}$X$_3$Y unit cell, (c) Example electron-localization-function (ELF) of LuH$_{11}$ at an isovalue of 0.55, (d) PDOS of the example LuH$_{11}$ showing metallic hydrogen states dominant at E$_F$.}
   \label{fig:LuH11XY}
 \end{figure}

 We present first a simple model structure for quaternary hydrides, with the chemical formula RH$_{11}$X$_3$Y (Fig. \ref{fig:LuH11XY}). It is distinguished by a central RH$_{11}$ unit with a typical electron localization function (ELF) and projected density of states (PDOS) illustrated using LuH$_{11}$ as an example in Fig. \ref{fig:LuH11XY}(c,d) surrounded by an X$_3$Y 'cage'. In the RH$_{11}$X$_3$Y model structure, the hydrogens are placed in the positions they occupy in FCC RH$_3$, except for the 1a Wyckoff position which may be occupied by hydrogen or a Lewis base (N,P,O,S). The 3d Wyckoff positions are also kept unspecified, but the intent for these atoms is to stabilize the FCC lattice. Choices of X and Y are also intended to make the X$_3$Y cage inert such the chemical bonding of the central RH$_{11}$ unit (and therefore its superconducting properties) is preserved. We explore the effects of filling the X and Y positions and hole doping in Fig. S5 (see SI), finding that the addition of trivalent X atoms typically leads to strong H bonding, bringing the hydrogen states far below E$_F$. Hole-doping and filling Y=O both result in hydrogen states dominating the DOS(E$_F$), though not at the same magnitude as LuH$_{11}$ [Fig. \ref{fig:LuH11XY}(d)]. These structures are simple enough to perform Eliashberg calculations for T$_c$, unlike the more complex structures discussed later. For the latter, estimate the critical temperature for hydride superconductors based on electronic indicators \cite{belli2021strong}, which we refer to as T$_c^{net}$.

  We first note that the networking-value T$^{net}_c$ [\eqref{eq:tc}] was found to be within $\pm 60$ K of isotropic Eliashberg T$^{Eli}_c$ calculations for 300 hydrides \cite{belli2021strong}. Calculations since then have verified its accuracy at predicting Eliashberg T$_c$s for CaH$_6$ (T$_{c}^{Eli}$: 236 K, T$_{c}^{net}$: 264 K) \cite{jeon2022electron} and Lu$_4$H$_{11}$N (T$_{c}^{Eli}$: 100 K, T$_{c}^{net}$: 99 K) \cite{fang2023assessing}. As a check on the T$_c$ estimation process, we estimate \eqref{eq:tc} for Fm$\overline{3}$m LuH$_{10}$ (see Fig. S4) at 150 GPa ($a = 4.91$ \AA) gives T$^{net}_c \approx 339$ K compared to the Eliashberg T$^{Eli}_c = 289$ K \color{blue}($\lambda \approx 3.5$)\color{black} for Fm$\overline{3}$m LuH$_{10}$ at 175 GPa \cite{fang2023assessing}. Additionally, our LuH$_{11}$ unit in Fig. \ref{fig:LuH11XY} ($a = 5.21$ \AA) is estimated to have a T$^{net}_c$ of 266 K. Our hypothesis is that any structure with similar chemical tendencies as these FCC RH$_{11}$ units would have a T$_c$ in a similar range. We note the similarity to the T$_c$ range reported experimentally for Lu-H-N \cite{salke2023evidence}. 
 
\subsubsection*{Lu$_4$H$_{11}$O Example}

 \begin{figure*}[t!]
   \centering
   \includegraphics[width=17.8cm]{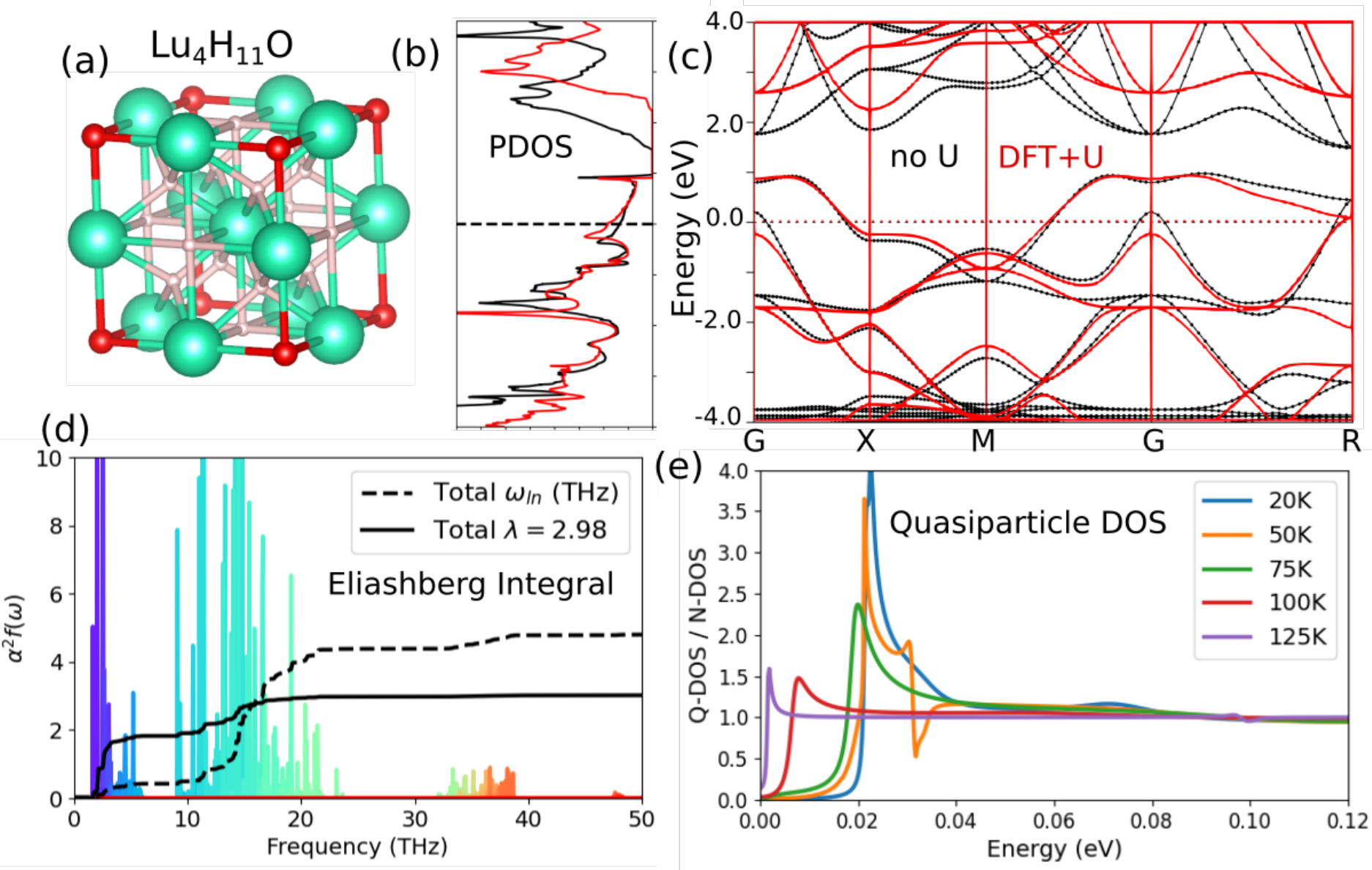}
   \caption{Electronic, structural, and superconducting properties of Pm3m Lu$_4$H$_{11}$O. (a) Visualization of the unit cell. (b) PDOS at 17 GPa with and without +U. (c) Band structures with and without +U. (d) $\alpha^2 F(\omega)$, integrated $\lambda(\omega)$, and integrated $\omega_{ln(\omega)}$ after convergence studies. (e) Quasiparticle density of states (QDOS) relative to the normal-state DOS. }
   \label{fig:Lu4H11O_dive}
 \end{figure*}

 As a reference for structural, electronic and Eliashberg calculations, we choose the example Pm$\overline{3}$m Lu$_4$H$_{11}$O (Fig. \ref{fig:Lu4H11O_dive}). The calculated DOS of the structure near E$_F$ is dominated by hydrogen states [DOS$_H$(E$_F$) $\sim 0.5$], meaning its predicted T$_c$ should be higher than that of Lu$_4$H$_{11}$N \cite{fang2023assessing} [DOS$_H$(E$_F$) $\sim 0.33$]. This arises from the difference in ionic bonding strength of O compared to N. We find through the ELF that the $\phi^{crit}_{net} = 0.521$, with the networking value predicted T$^{net}_c = 130$ K. Pm$\overline{3}$m Lu$_4$H$_{11}$O features a nearly-flat electron band 0.36 eV below E$_F$ (around $X$) [Fig. \ref{fig:Lu4H11O_dive}(c)] which is expected to contribute to the superconductivity due to the high energy of phonons around 12-15 THz (about 0.08 eV) and 35 THz (0.15 eV). We find increasing correlation effects via DFT+U (U$_d$ = 5.5 on Lu) can induce Lifshitz transitions [Fig. \ref{fig:Lu4H11O_dive}(c)] indicating correlation effects could increase the EPC. Density functional perturbation theory calculations on a 4x4x4 q-point mesh indicate it is dynamically stable above 25 GPa. We used the EPW code \cite{ponce2016epw} to interpolate the electron-phonon matrix elements using Wannier functions, with k/q meshes of 32$^3$ and 16$^3$. The $\alpha^2 F(\omega)$ leads to a large integrated electron-phonon coupling of $\lambda \approx 3$ [Fig. \ref{fig:Lu4H11O_dive}(d)] and $\omega_{log} \approx 5$ THz, with an Allen-Dynes T$_c$ (see Eq. \ref{eq:allendynes}) of 90 K using $\mu = 0.10$ \cite{allen1975superconductivity}. The Allen-Dynes T$_c$ decreases to 50 K if not converged (e.g. using a coarser mesh and larger thermal broadening factor). The T$_c$ has sizable contributions from the octahedral hydrogen modes below 5 THz and 12-15 THz, with a small contribution to $\omega_{log}$ from higher-frequency tetrahedral hydrogen modes at 35 THz. 

 We performed anisotropic Eliashberg calculations as implemented in the EPW software \cite{margine2013anisotropic} and compute the quasiparticle DOS in the superconducting state at various temperatures, illustrating the gap closure around 125 K [Fig. \ref{fig:Lu4H11O_dive}(g)] which coincides with the T$^{net}_c$ of 130 K predicted by \eqref{eq:tc}. The sizable difference between the Allen-Dynes and Eliashberg T$_c$ may reflect that the Allen-Dynes formula was created from fitting to alloys of conventional superconductors \cite{allen1975superconductivity} with lower phonon frequencies, and so the high frequency modes here (15 THz and 35 THz) may not be appropriately captured in these fits. The phonon dispersion, superconducting properties, and effect of doping are further elucidated in the Supplementary Information. 

\subsubsection*{Other RH$_{11}$X$_3$Y Examples}
 
Both LuH$_{11}$Ca$_3$O and CaH$_{11}$Y$_3$O have dominant DOS$_H$(E$_F$) and end up having similar T$_c^{net}$ despite the almost threefold difference in their total DOS$_{tot}$(E$_F$) (Fig. S6, SI). We emphasize this is due to the observation that T$_c$ is not correlated with the total DOS(E$_F$) in hydrides, but the relative hydrogen contribution \cite{belli2021strong}. We also 

As boron is known to attain an oxidation state as low as B$^{5-}$, we also explored Lu$_4$H$_{11}$B (Fig. S7, SI) and found an interesting flat band near E$_F$ of Lu-d, H-s, and B-p character which remained under relaxation that would likely lead to an instability of some kind. We also found SiH$_{11}$Lu$_3$N to have a hybridized nearly-flat band near E$_F$ (Fig. S22, SI) with T$_c^{net}$ = 90 K. As their predicted T$^{net}_c$ are below 200 K we did not further examine these examples. We confirm there is no DFT+U dependence for these results (the Fermi level properties do not change).

\begin{table*}[t!]
  \centering
  \footnotesize
  \caption{Table of selected materials, with the parent structure, computed electronic properties, and estimated$^a$ T$_c$}
\begin{tabular}{llrrrcc}
Parent Structure & Material & a (\AA) & DOS$_{H}$(E$_F$) & $\phi^{crit}_{net}$ & T$_c^{net}$ (K) & Notes \\
\midrule
  RH$_{11}$X$_3$Y & Lu$_4$H$_{11}$O & 5.01 & 0.66 & 0.50 & 130 & T$_c^{Eli}$ = 125 K with frequency-dependent s-wave gap \\
                 & LuH$_{11}$Ca$_3$O & 5.11 & 0.62 & 0.52 & 145 &  \\
                 & CaH$_{11}$Y$_3$O & 5.14 & 0.66 & 0.485 & 130 &  \\
                 & ScH$_{11}$Y$_3$N & 5.05 & 0.5 & 0.495 & 120 &  \\
                 & SiH$_{11}$Lu$_3$N & 4.90 & 0.4 & 0.47 & 90 & T$_c^{net}$ rises to 125 K with a supercell distortion \\    
\midrule  
  (RH$_{11}$)$_2$X$_6$YZ & Lu$_8$H$_{23}$N & 10.04 & 0.74 & 0.515 & 165-180 & Dynamically stable at 2 GPa due to quantum effects \\  
                 & Lu$_8$H$_{22}$N & 10.07 & 0.7 & 0.52 & 160 & T$_c^{net}$ = 215 K with classical distortion and doping \\ 
                 & Sc$_6$Ca$_2$H$_{23}$N & 9.60 & 0.5 & 0.582 & 160 & T$_c^{net}$ = 220 K with classical distortion \\  
                 & Y$_6$Mg$_2$H$_{23}$O & 10.07 & 0.8 & 0.50 & 165-170 & T$_c^{net}$ = 170 K by removing 4b-H \\
                 & Y$_6$Ca$_2$H$_{22}$N (U$_{Y_d}$ = 3.5 eV) & 10.32 & 0.85 & 0.51 & 165 & Removed 4b-H\\  
                 & Mg$_6$La$_2$H$_{23}$O & 10.05 & 0.77 & 0.575 & 200 & Most likely unstable \\
                 & Mg$_6$Sc$_2$H$_{23}$O & 9.50 & 0.77 & 0.575 & 200 & Most likely unstable \\
\midrule    
  (R$^1$H$_{11}$)(R$^2$H$_{11}$)X$_6$YZ & Y$_7$LiH$_{23}$N & 10.18 & 0.66 & 0.55 & 170 & Hydrogen sublattice may distort, remains metallic \\                 
                 & Y$_7$LiH$_{23}$N (U$_{Y_d}$ = 3.5 eV) & 10.18 & 0.79 & 0.54 & 185-220 & T$_c^{net}$ = 220 K with electron doping \\
                 & Y$_7$LiH$_{22}$N (removed 4b-H) & 10.22 & 0.66 & 0.61 & 190-220 & T$_c^{net}$ = 220 K with electron doping \\
                 & Y$_7$CaH$_{23}$N & 10.24 & 0.65 & 0.56 & 180-200 & T$_c^{net}$ = 200 K with electron doping \\
                 &  Lu$_7$SiH$_{23}$N & 9.96 & 0.5 & 0.55 & 150 & \\
                 & Sc$_7$LiH$_{23}$N & 9.33 & 0.48 & 0.655 & 190 & ScH$_3$-based compounds may need 20-40 GPa \\
                 & Sc$_7$LiH$_{23}$N (U$_{Sc_d}$ = 5 eV) & 9.33 & 0.51 & 0.65 & 195 & \\  
                 & Sc$_7$NaH$_{23}$N & 9.43 & 0.51 & 0.63 & 185 & \\
                 & Sc$_7$NaH$_{23}$N (U$_{Sc_d}$ = 5 eV) & 9.43 & 0.64 & 0.62 & 205 & \\
                 & Sc$_7$MgH$_{23}$N (U$_{Sc_d}$ = 5 eV) & 9.43 & 0.7 & 0.64 & 220-230 & T$_c^{net}$ = 230 with electron doping \\  
  \bottomrule  
\end{tabular}

\caption*{$^a$ The estimated T$_c^{net}$ is rounded to the nearest 5 K with an error bound of $\pm 60$ K \cite{belli2021strong}.}
\label{tab:examples}
\end{table*}

Selected examples discussed so far as well as the others we explore later are outlined in Table \ref{tab:examples}, with some relative energies of formation computed in Table S1. The results establish that while the first model structure (Fig. \ref{fig:LuH11XY}) can result in hydrogen-dominant states at E$_F$, the magnitude of the DOS(E$_F$) is not comparable to that of the core RH$_{11}$ structure; the DOS$_{H,rel}$(E$_F$) ratio is not typically close to 1, which limits the achievable T$^{net}_c$ (Eq. \ref{eq:tc}). 
 
\subsection*{Fm$\overline{3}$m (RH$_{11}$)$_2$X$_6$YZ Structure}

 \label{sec:RH11}

\begin{figure}[t!]
   \centering
     \includegraphics[width=1.0\linewidth]{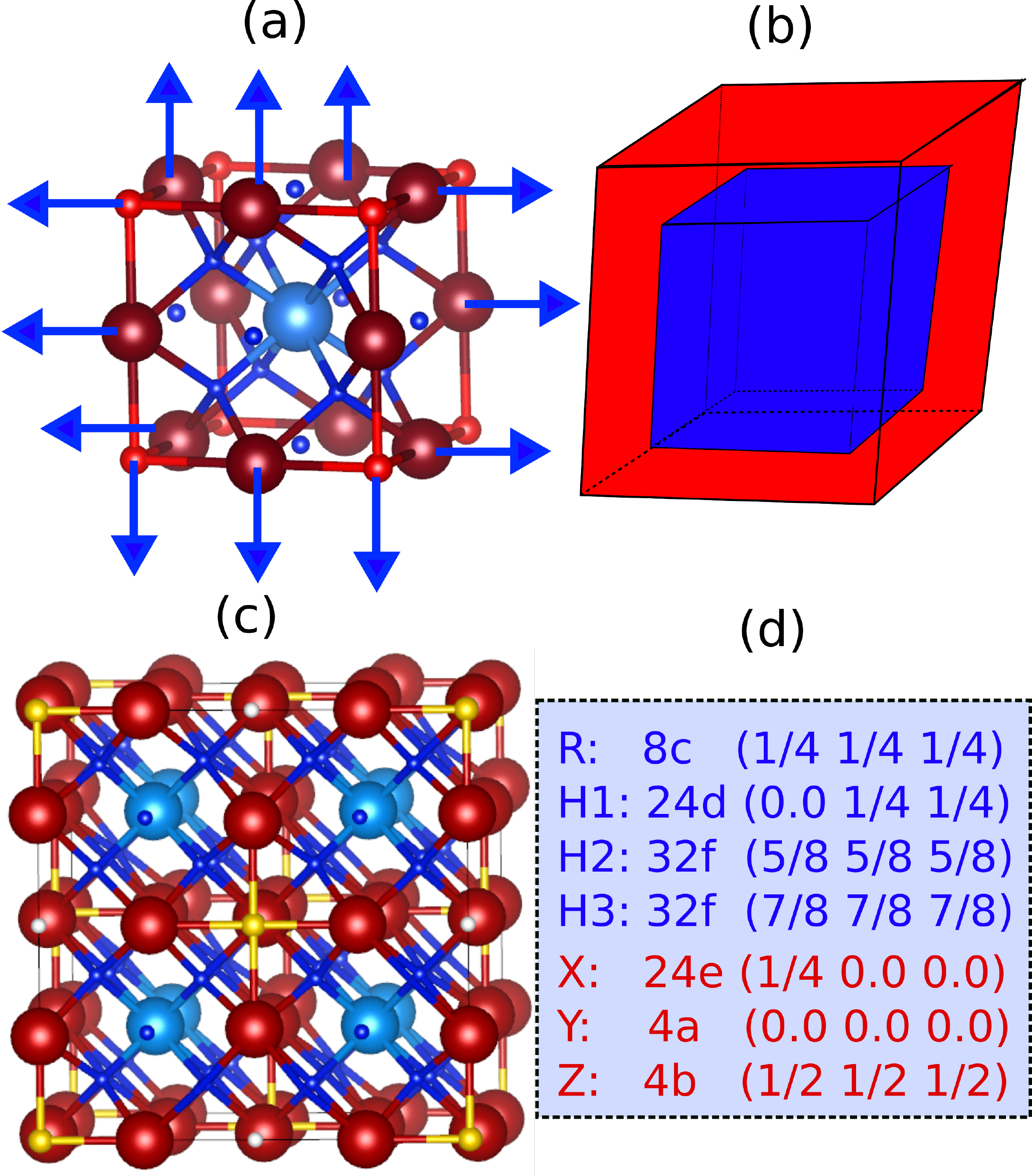}
   \caption{ Strategy for creating the (RH$_{11}$)$_2$X$_6$YZ structure. In this scheme we redirect the charge of X$_3$Y layers of the simple model structure (Fig. \ref{fig:LuH11XY}) away from the RH$_{11}$ units. (a) Visualization of extra charge brought by 3d-X atoms being redirected away from the RH$_{11}$ unit. (b) A depiction of an inert X$_6$YZ shell containing the RH$_{11}$ unit. (c) Plot of (RH$_{11}$)$_2$X$_6$YZ using a similar coloring scheme as (b) to further illustrate the concept. (d) The Wyckoff positions of the quaternary structure. The atoms in (c) are colored as: 24e-X (red), 4a-Y (yellow), 4b-Z (white), and RH$_{11}$ units (blue).}
   \label{fig:strat}
 \end{figure}

 Using T$_c^{net}$ [\eqref{eq:tc}] as a guiding principle, we build on the simple RH$_{11}$X$_3$Y structure (Fig. \ref{fig:LuH11XY}) and design a supercell to increase H$_f$ and $\text{DOS}_H(E_F)$. The increased H$_f$ requires less [O,N,S,P], and we hypothesize that high symmetry Fm$\overline{3}$m structures inspired by our previous work \cite{denchfield2024electronic, denchfield2024quantum} will lead to large $\text{DOS}_H(E_F)$ by bringing tetrahedral hydrogen states to E$_F$. This leads to a supercell in which charge is redirected away from the X$_3$Y cage surrounding the RH$_{11}$ unit [Fig. \ref{fig:strat}(a-b)]. We visualize and specify a model structure meeting these requirements in Fig. \ref{fig:strat}(c,d). The blue regions indicate the RH$_{11}$ regions. 
 

 
 Our second model structure (Fig. \ref{fig:strat}) is designed to emulate the hydrogen-dominated band structure near E$_F$ of the core RH$_{11}$ units, with the appropriate choice of X,Y, and Z atoms. The chemical formula is (RH$_{11}$)$_2$X$_6$YZ, where R is an atom in the 8c position, X is an atom stabilizing the FCC RH$_{11}$ lattice in the 24e position with $x \approx 0.25$, and Y,Z in the 4a/4b positions ensure the X atoms are inert with respect to the RH$_{11}$ lattice. The hydrogens occupy two 32f positions with $x \approx 0.875$ and $x \approx 0.625$ as well as the 24d positions. Figure S12 shows four examples without relaxing the structure from the parameters given in Fig. \ref{fig:strat}(d). Their electronic properties suggest a high estimated T$_c$ in the 150-170 K range using \eqref{eq:tc}. However, these examples are unrelaxed so we do not quantify their properties further.


\subsubsection*{(CaH$_{11}$)$_2$ Sc$_6$NH Example}

 \begin{figure}
   \centering
     \includegraphics[width=1.0\linewidth]{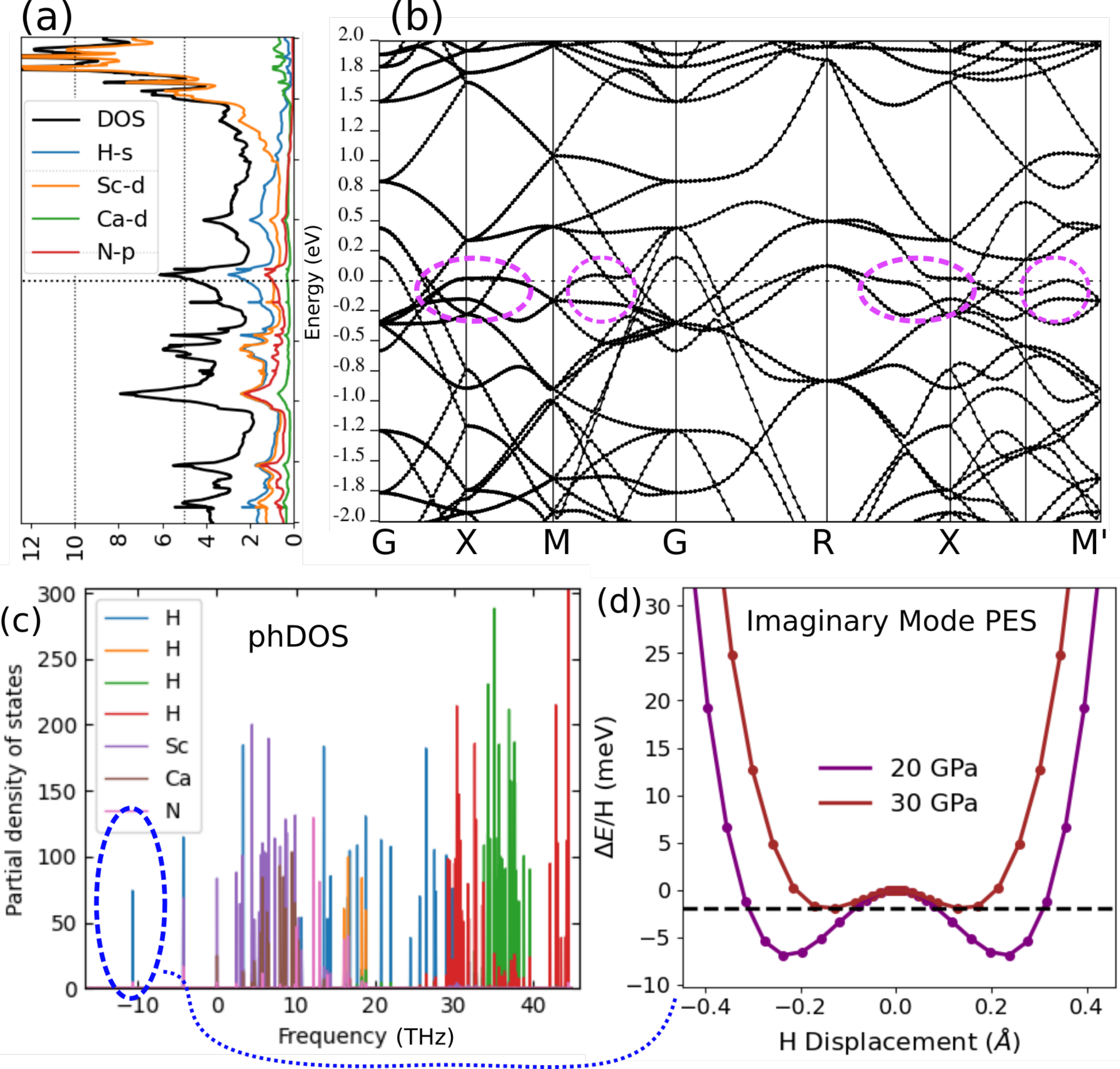}
   \caption{Electronic density of states (a) and band structure (b) of Sc$_6$Ca$_2$H$_{23}$N at 20 GPa, with several van Hove singularities near E$_F$ circled, (c) Phonon density of states resolved by atomic projections of unique Wyckoff positions, (d) Potential energy surface (PES) of the mode with largest imaginary frequency at 20 and 30 GPa.}
   \label{fig:Sc6Ca2H23N}
 \end{figure} 

We start with an in-depth look at a particular example of the (RH$_{11}$)$_2$X$_6$YZ model structure. Choosing R=Ca, X=Sc, Y=N, and Z=H, we have (CaH$_{11}$)$_{2}$Sc$_6$NH, or Sc$_{0.75}$Ca$_{0.25}$H$_{2.875}$N$_{0.125}$. We focus on this example for a variety of reasons. In terms of stability, 25-33\% Ca doping is predicted to decrease the pressure needed to stabilize cubic ScH$_{3}$ by over 10 GPa \cite{villa2022superconductivity}. With respect to its potential for high-T$_c$ superconductivity, CaH$_{6}$ has been shown to exhibit a T$_c$ of 215 K \cite{ma2022high} close to the original prediction \cite{wang2012superconductive} and possibly higher T$_c$ with more hydrogen-rich compounds \cite{shao2019unique}. Additionally, doping CaH$_6$ with rare earths could achieve even higher T$_c$ \cite{yang2022stable, papaconstantopoulos2023high}. Finally, these ingredients the following criteria: ScN is semiconducting, the Sc atoms will help stabilize the tetrahedral hydrogen in the CaH$_{11}$ unit, and the large electonegativity of N is expected to weaken the ionic bonding of Sc atoms with tetrahedral hydrogen, bringing tetrahedral hydrogen states higher in energy to hybridize with the octahedral hydrogen.

The electronic and phonon properties of (CaH$_{11}$)$_{2}$Sc$_6$NH at 20 GPa are shown in Fig. \ref{fig:Sc6Ca2H23N}. The PDOS/bandstructure shows multiple hydrogen-dominant van Hove singularities (vHs) near E$_F$, reminiscent of the flat bands in Lu$_8$H$_{23}$N \cite{denchfield2024electronic, denchfield2024quantum}. We find $\phi^{crit}_{net} = 0.585$ and T$^{net}_c = $ 165 K. Removing the 4b-H or replacing the nitrogen with oxygen both result in T$_c^{net} $ rising to 180 K (Fig. S15). The imaginary phonon frequencies at $\Gamma$ indicate there may be a structural instability of the hydrogen sublattice.

To evaluate the likelihood of quantum effects stabilizing this structure against such an instability, we evaluate the anharmonicity of the potential energy surface (PES) associated with displacing the crystal along the corresponding dynamical matrix eigenvector (visualized in the inset). We define the number of hydrogen strongly displaced in the mode to be all hydrogen whose displaced distance is more than $D_{max}/6$, where D$_{max}$ is the maximum distance a hydrogen is displaced. We see the maximum energy gain at 20 (30) GPa is 7 (2) meV/hydrogen at an average hydrogen displacement of 0.22 (0.15) \AA. The double well PES becomes more harmonic with increased pressure, but the wells deepen as pressure is decreased, similar to Lu$_8$H$_{23}$N \cite{denchfield2024quantum}. Fig. S13 shows examples of ground state densities computed for anharmonic potentials; potentials can be surprisingly anharmonic while still favoring a centered wavefunction. These results indicate that nuclear quantum effects stabilize the structure against this instability at this or a slightly higher pressure. For pressure ranges where this distortion occurs, we find this distortion in fact could enhance T$_c^{net}$ up to 220 K based on our calculations at 20 GPa (Fig. S2). For more information on this and related stoichiometries, see the Supplemental Information.

\subsubsection*{Other (RH$_{11}$)$_2$X$_6$YZ Examples}

The electronic properties of twelve more examples are shown in Fig. S9. All twelve examples have hydrogen states dominant at E$_F$, which correlates with high T$_c$ hydride superconductivity \cite{belli2021strong}. As H$_f$ in \eqref{eq:Hf} is nearly constant in these examples, the only remaining parameter to maximize T$_c$ is $\phi^{net}_{iso}$. Using this metric, we find the highest $\phi^{net}_{iso}$ values in the (RH$_{11}$)$_2$(Mg$_6$OH) systems, with a corresponding networking-value T$_c^{net}$ of 200 K. For R=La the high $\phi^{net}_{iso}$ responsible for this is attributed to hydrogens tetrahedrally oriented with respect to Mg bending away from the oxygen atoms and forming a body-centered cubic hydrogen cage (Fig. S20), with H-H distance 1.964\AA \ between cube edges and 1.692\AA \ between the vertices and the central H atom. Such small H-H distances have also been observed in some intermetallic hydrides \cite{klein2022neutron}. For R=Sc, the hydrogen cubic cage does not form, but the smaller lattice constant appears to make up for it in terms of high $\phi^{net}_{iso}$ values. Unfortunately, a stability analysis (see SI Section ``Stability'') indicates ([La,Sc]H$_{11}$)$_2$Mg$_6$OH are dynamically unstable. However, we note the ternary system (LuH$_{11}$)$_{2}$Lu$_6$HN was predicted to be dynamically stable at 2 GPa when considering quantum effects \cite{denchfield2024quantum}. Clearly, fine-tuning of the stoichiometry can be a determining factor in each structure's dynamical stability. 

 We find that structures with dominant hydrogen states at E$_F$ can be obtained by choosing X=Sc/Y/La/Lu, R = Li/Na/Mg/Ca/Sr/Al/Ga/In, Y = H/O/S/N/P, and Z = H or vacant (see SI). Swapping the role of the R and X atoms also results in hydrogen-dominant states at E$_F$ (see SI), though dynamical and thermodynamic stability is in question. 

The properties of a few (RH$_{11}$)$_2$X$_6$YZ examples are summarized in Table \ref{tab:examples}. While T$_c^{net}$ is generally higher for these examples the RH$_{11}$X$_3$Y model structure, we find in (CaH$_{11}$)$_2$ Sc$_6$NH that relaxation along instability pathways can in fact raise T$_c^{net}$ through large increases in $\phi^{net}_{iso}$ (Fig. S2). A similar vibrational analysis on (LuH$_{11}$)$_2$ Lu$_6$N also reveals a classical distortion \cite{denchfield2024quantum}, which coupled with fine-tuning of the doping level, raises T$_c^{net}$ to 215 K. Based on these observations, we propose our last model structure, designed to increase $\phi^{net}_{iso}$ and therefore T$_c^{net}$. 

\subsection*{F$\overline{4}$3m (R$^1$H$_{11}$)(R$^2$H$_{11}$)X$_6$YZ Structure}

This structure involves symmetry breaking of the 8c-R atoms of the previous model structure, captured with the formula (R$^{1}$H$_{11}$)(R$^2$H$_{11}$)(X$_6$YZ), where X is a rare earth atom, and R$^1$, R$^2$ are other electron-donating metal atoms (possibly the same as X). The Wyckoff positions are summarized in Fig. \ref{fig:threestructs}. The change in metal ion size enhances symmetry breaking of the 24d-H positions in the structure. Those hydrogen atoms are observed to shift to either R$^1$ or R$^2$ (typically whichever has more valence electrons). Fig. \ref{fig:threestructs} shows the symmetry breaking hydrogens (red) as closer towards the 4d-R$^2$ atoms.

We use this generalization of the Fm$\overline{3}$m (RH$_{11}$)$_2$X$_6$YZ to simulate 12.5\% doping of ternary R$_8$H$_{23}$N structures, which involves choosing X = R$^2$ = a rare earth, and R$^1$ a dopant atom. 

\subsubsection*{Doping R$_8$H$_{23}$N}

 \begin{figure}
   \centering
     \includegraphics[width=1.0\linewidth]{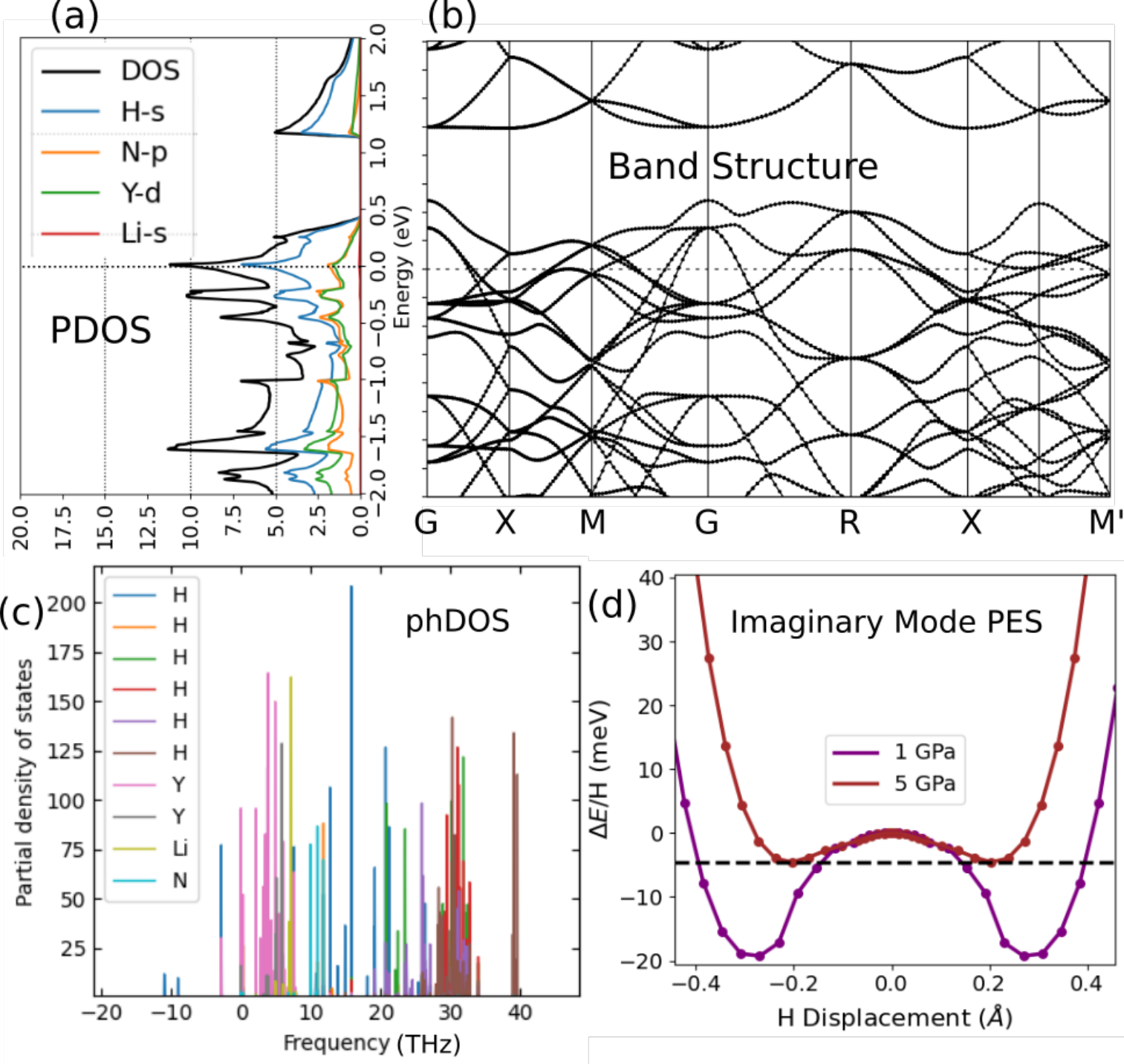}
   \caption{Properties of (YLiH$_{22}$)(Y$_6$NH) (a = 10.1 \AA) (a) Electronic band structure and projected density of states, (b) Phonon density of states resolved by atomic projections of unique Wyckoff positions, (c) Potential energy surface (PES) of the mode with largest imaginary frequency at 1 GPa and 5 GPa. }
   \label{fig:Y7LiH23N}
 \end{figure} 

We begin with the interesting electronic structure of Lu$_8$H$_{23}$N (A) \cite{denchfield2024electronic} and investigate replacing Lu with Sc,Y. Cell relaxation was performed and the resulting PDOS and electronic properties are compared in Fig. S19 (see SI). U-values were chosen by performing DFT+U calculations on the rare earth nitrides and adjusting U$_d$ to match the indirect band gaps to experiment \cite{liu2014electronic}. For Y$_8$H$_{23}$N, while the flat hydrogen band is still present near E$_F$, the larger lattice constant results in a smaller $\phi^{net}_{iso}$, lowering the estimated T$_c^{net}$. For Sc$_8$H$_{23}$N the much smaller lattice constant leads to a higher $\phi^{net}_{iso}$ but the Sc$_d$ orbitals hybridizing with the Fermi level leads to more charge screening of the negative nitrogen atoms. This brings hydrogen states below E$_F$ and weakens the flatness of the hydrogen band. 

Given that doping YH$_3$ assists in stabilizing the FCC phase \cite{van2001insulating, kataoka2019facile, kataoka2021zirconium, CARSTEANU200582} and various studies indicate doping rare-earth hydrides yields high T$_c$s \cite{villa2022superconductivity, sun2022high, liang2019potential}, we investigate the effects of substituting some R=[Sc,Y] atoms in R$_8$H$_{23}$N with alkali and alkali earth atoms. In particular, we investigate Li, Na, and Mg doping in Sc$_8$H$_{23}$N, and Li, Mg, and Ca doping in Y$_8$H$_{23}$N. We start with the parent structure in Fig. \ref{fig:threestructs}(right) using X=Sc/Y,R$^1$=D (dopant), R$^2$=Sc/Y, Y=H,Z=N, with the R$^1$ position occupied by either a dopant D=Li/Mg/Ca or Sc/Y for lower dopant concentrations.

\subsubsection*{(YH$_{11}$)(LiH$_{11}$)Y$_6$HN Example}

A representative example of the F$\overline{4}$3m model structure [Fig. \ref{fig:threestructs}(right)] is (YH$_{11}$)(LiH$_{11}$)Y$_6$HN (or Y$_{0.875}$Li$_{0.125}$H$_{2.875}$N$_{0.125}$), which has a lattice constant of 10.1 \AA\ whose stoichiometry is similar to an experimentally stable cubic structure Y$_{0.9}$Li$_{0.1}$H$_{2.8}$ \cite{kataoka2018stabilization}. Figure \ref{fig:Y7LiH23N} illustrates its electronic and vibrational properties via DFT and frozen phonon calculations \cite{togo2023first}. As in previous examples, the DOS$_H$(E$_F$) is high and dominant [Fig. \ref{fig:Y7LiH23N}(a)], with contributions from all hydrogen, though the largest contribution is from the 4b-H. The band structure illustrates a propensity of vHs near E$_F$, as well as flat band regions 0.3 eV below and 1.1 eV above E$_F$ [Fig. \ref{fig:Y7LiH23N}(b)]. As designed, the symmetry-broken octahedral hydrogen improve the connectivity of the hydrogen sublattice, increasing $\phi^{net}_{iso}$ to 0.6 (note our record of $\phi^{net}_{iso}$ for Y-based hydrides using the previous structure was 0.55; see Fig. S9). This increases T$_c^{net}$ to 185 K, which may be increased by pressure and further fine-tuning of the stoichiometry. We find electron doping can increase T$_c^{net}$ up to 220 K. 

Since the zero pressure phonon calculation reveals imaginary modes [Fig. \ref{fig:Y7LiH23N}(c)], we perform an anharmonicity analysis as was done previously for (CaH$_{11}$)$_2$Sc$_6$NH (Fig. \ref{fig:Sc6Ca2H23N}). Here we find another double well, which at 1 GPa is quite deep but becomes nearly flat at 5 GPa [Fig. \ref{fig:Y7LiH23N}(d)]. All frequencies are positive at 15 GPa, implying the high-symmetry F$\overline{4}$3m structure becomes dynamically stable in the 5-15 GPa range. The distortion expected to occur below 5 GPa features octahedral hydrogen bending towards tetrahedral hydrogen, which is expected to anisotropically change $\phi^{net}_{iso}$. We find the distortion at 1 GPa to have a fairly negligible effect on the PDOS like in (CaH$_{11}$)$_2$Sc$_6$NH, but the T$_c^{net}$ drops to 145 K due to an undesirable anisotropy in $\phi^{crit}_{net}$.

Various perturbations on (YH$_{11}$)(LiH$_{11}$)Y$_6$HN also seem to yield high T$_c^{net}$. Removing the 4b-H yields F$\overline{4}$3m Y$_{0.875}$Li$_{0.125}$H$_{2.75}$N$_{0.125}$ with T$_c^{net} = 200$ K and a similar double well PES for the mode with largest imaginary frequency (Fig. S3). The primary difference from removing the 4b-H is the DOS is fairly flat around E$_F$, but an extremely flat hydrogen-based band exists 1 eV above E$_F$, corresponding to 'hydrogen-cage-like' nearly-localized electronic states. Further changing one of the 4a-N to a hydrogen atom effectively electron-dopes the system, which also has a high T$_c^{net}$ of 200 K after relaxation. 

\subsubsection*{Other (R$^1$H$_{11}$)(R$^2$H$_{11}$)X$_6$YZ Examples}

(YH$_{11}$)(CaH$_{11}$)Y$_6$HN has a comparatively modest $\phi^{net}_{iso}$ of 0.56 and T$^{net}_c$ of 180 K [Fig. S16]. We proceed with Sc-based examples, as the reduced lattice constant generally favors high $\phi^{net}_{iso}$. Our remaining examples involve doping Sc$_8$H$_{23}$N to form F$\overline{4}$3m Sc$_7$[Li,Na,Mg]H$_{23}$N. The electronic and lattice parameters, and T$_c$ estimates are shown in Table \ref{tab:examples} with PDOS plots in the SI. Behavior similar to Li-doped Y$_8$H$_{23}$N occurs with H shifting towards the 4d-Sc Wyckoff positions. This change coupled with the smaller lattice constant of the FCC ScH$_3$ parent structure leads to large $\phi^{net}_{iso}$ due to the robust hydrogen network that forms around the 4d-Sc positions.

We summarize the properties of these examples in Table \ref{tab:examples}. The T$_c$ estimates of 185-230 K for these structures illustrate the power of combining both nitrogen and alkali metal dopants in this structure design, enabling heightened $\phi^{net}_{iso}$. We mention we briefly investigated replacing N with S, and found F$\overline{4}$3m Sc$_7$MgH$_{23}$S is insulating but with a high $\phi^{net}_{iso}$ = 0.655. Hole-doping, for example with further Mg (perhaps using Fm3m (MgH$_{11}$)$_2$Sc$_6$SH) would create a metal with hydrogen-dominant states.
 
\subsection*{Discussion}
\label{sec:conc}

In the search for hydride superconductors with very high T$_c$, a complicating factor is the necessity to include quantum zero-point motion/energy and thermal effects. Without such effects, harmonic phonons with classical nuclei often overestimate structural instabilities for high T$_c$ hydrides \cite{liu2018dynamics, errea2020quantum}. In fact, T$_c$ tends to be highest when the system is closest to structural transitions \cite{quan2019compressed} (and where DFT methods can fail). Another complicating factor for first principles searches is that of metastability; it may prove sufficient for experimental purposes to produce a metastable superconductor (so long as its lifetime is measured in months or years), yet such structures would be discarded in many computational structure searches. We therefore take a different approach and design complex hydride structures with the intent of producing electronic properties that correlate with high T$_c$ \cite{belli2021strong}. We use FCC RH$_3$ as the parent structure, as these structures have high EPC and can be stabilized at ambient pressures in a variety of ways \cite{kataoka2021face, kataoka2019facile, kataoka2022origin,  villa2022superconductivity, li2024stabilization}. 

We have proposed three successively more chemically complex model hydride structures with predicted T$^{net}_c$ maximized by having hydrogen-dominant states at E$_F$ and a strongly connected hydrogen sublattice. The three models exhibit increased flexibility for tuning stability and T$_c$ relative to simple binary structures. Examples within each model structure are candidate high T$_c$ superconductors that are potentially stable near ambient pressure and therefore deserve additional study. 

Compounds based on the first and simplest model structure show moderate promise in producing ternary/quaternary hydride superconductors with T$_c$ in the 100-150 K range and are potentially metastable near ambient pressures. We find Pm$\overline{3}$m Lu$_4$H$_{11}$O has a lower formation energy than LuH$_3$ + 3LuH$_2$ + H$_2$O by $0.128$ eV/atom (Table S1). We also note YH$_3$ and Y$_2$O$_3$ are known to react to form cubic YH$_x$O$_y$ under heating \cite{zapp2019yho}. The second and third model structures of quaternary hydrides have enough flexibility to optimize parameters going into the T$^{net}_c$ estimator, which we use to generate examples with upper estimated T$^{net}_c$s up to 220-230 K at near-ambient pressures when accounting for hydrogen distortions. While the examples discussed are not on the convex hull (Table S1), this does not discount their possible metastability. For example, diamond is also 0.14 eV/atom above its convex hull using data from Ref. \cite{jain2020materials}. 

The second model can produce compounds with high T$_c^{net}$ stable at low pressures, as Lu$_8$H$_{23}$N (A) studied by us previously \cite{denchfield2024electronic} was shown to be dynamically stable at 2 GPa when nuclear quantum effects are included \cite{denchfield2024quantum}. Recently, N-doped FCC LuH$_3$ has been observed to be recoverable to ambient \cite{li2024stabilization} using the techniques in Ref. \cite{kataoka2022origin}. Our Fm$\overline{3}$m (CaH$_{11}$)$_2$ Sc$_6$NH example is expected to be dynamically stable above 20 GPa when nuclear quantum effects are included. As the double well is deepened (from i.e. lower pressure) the octahedral hydrogen are expected to distort from high symmetry positions which surprisingly increases T$_c^{net}$, with a pressure regime where spectroscopic signatures are broadened due to quantum ZPM. This interpretation on the structural dynamics is supported by high-pressure studies of ScH$_3$ \cite{kume2011high}.

Our third model structure involves complexity higher than quaternary systems. Nevertheless, we limit ourselves to the quaternary case by analyzing the doping of [Sc/Y]$_8$H$_{23}$N structures. The structure is designed to distort hydrogen to increase $\phi^{net}_{iso}$ with many possible dopants. Based on our calculations on cubic Y$_{0.875}$Li$_{0.125}$H$_{2.875-\delta}$N$_{0.125-\epsilon}$ structures, we predict T$_c$ to have a maximum of 220 K $\pm$ 60 K at a pressure between 5-15 GPa for optimal $\delta, \epsilon$. For synthesis we recommend leveraging the approaches used in Refs. \cite{kataoka2018stabilization, kataoka2022origin, li2024stabilization}. At lower pressures, distortions away from the F$\overline{4}$3m structure occur. It is difficult to predict what the effect on T$_c$ is because the $\phi^{net}_{crit}$ parameter becomes very anisotropic; small four-center hydrogen bond networks form for  with $\phi^{*,net}_{crit} \equiv 0.8$ but extended hydrogen networks form across the entire cell at $\phi^{net}_{crit} = 0.5$. Such anisotropic behavior in large unit cells was not considered in the data for T$_c^{net}$ \cite{belli2021strong}, so the predicted T$_c^{net}$ of 145 K for the distorted structure at 1 GPa could have a larger error bar than 60 K. Other examples of the F$\overline{4}$3m model structure similarly illustrate that doping R$_8$H$_{23}$N  increases T$_c^{net}$ by 30-50 K. As the highest predicted T$_c^{net}$ comes from cubic (Mg,N)-doped ScH$_3$, we note Mg and Sc form cubic alloys \cite{schob1965ab} and in fact have been studied as hydrogen storage materials \cite{kalisvaart2006electrochemical, al2024computational}. Based on our results and previous studies \cite{van2001insulating, kataoka2019facile,villa2022superconductivity}, we find it likely that these doped [Sc,Y]$_8$H$_{23-\delta}$[N,O] examples with extremely high predicted T$_c$ are synthesizable and dynamically stable with compression. 

We note again T$^{net}_c$ only reproduces T$_c$ from Eliashberg calculations within $\pm 60$ K, so some examples presented may in fact have T$_c$ exceeding 270 K. It is likely that the T$_c$ for the structures presented here is large in a narrow pressure range, based on Eliashberg calculations on YH$_3$ \cite{kim2009predicted} and experiments on N-doped LuH$_3$ \cite{dias2023observation, salke2023evidence}. We remark that preliminary constrained DFT calculations indicate strong correlation effects may be present in these systems for the same reasons as presented in Ref. \cite{eder1997kondo}; that is, DFT struggles to represent the near-neutral hydrogen atoms present in RH$_{3-\delta}$ and how its charge fluctuations impacts the electronic structure. A more thorough investigation of these effects is underway. 


In summary, we present model structures based on doped supercells of FCC RH$_3$ to adopt their dynamical stability properties at low pressures while tuning their electronic properties. These supercells have been engineered to have high T$_c$s with upper estimates of 220 $\pm 60$ K. We end by remarking that it is compelling that the highest predicted T$_c$ using our methodology comes from (Mg,N)-doped ScH$_3$, which features both Ashcroft's proposal for hydrogen-dominant metallic alloys \cite{ashcroft2004hydrogen} and Feynman's guess that the highest temperature superconductor would be Sc-based \cite{goodstein2000richard}. 

\section*{Materials and Methods}

\label{sec:methods}

  We use density functional theory (DFT) \cite{kohn1965self, giannozzi2009quantum, giannozzi2017advanced} and DFT+U \cite{wang2016local, tolba2018dft+} first-principles electronic structure calculations to study the examples built using our proposed model structures (Fig. \ref{fig:threestructs}). The SSSP database \cite{prandini2018precision} was used to vet pseudopotentials for convergence of the computed pressure, phonon frequencies, and formation energies. Each relaxed structure had remaining forces under \texttt{1e-5} Ry/Bohr, and energy differences were under \texttt{1e-5} Ry. We used wavefunction plane-wave cutoffs of 75 Ry for the structures and charge density cutoffs of 300 Ry, though we performed convergence tests with higher values (especially for Sc-based compounds) and found no differences. We generally used tight k-meshes corresponding to 0.13 \AA$^{-1}$, but did convergence tests with more k-points as needed. We used \texttt{Quantum Espresso}'s default atomic projections for the PDOS calculations. The \texttt{phonopy} software \cite{togo2023first} was used the compute phonon dispersions and generate the eigenvectors for atomic displacements. 

 The Allen-Dynes formula for T$_c$ takes the form 
 \begin{align}
   \label{eq:allendynes}
   &T_c \approx f_1 f_2 \frac{\omega_{ln}}{1.20} \exp{-1.04\qty(\frac{1+\lambda}{\lambda - \mu^* (1-0.62\lambda)})} \\
   &f_1 = \qty[1 + \qty(\frac{\lambda}{\Lambda_1})^{3/2}]^{1/3} \quad \quad f_2 = 1 + \frac{\lambda^2}{\lambda^2 + \Lambda_2^2} \qty(\frac{\overline{\omega}_2}{\omega_{ln}} - 1) \\
   &\Lambda_1 = 2.46 (1 + 3.8\mu^*) \quad \quad \quad   \Lambda_2 = 1.82 (1+6.3\mu^*)\frac{\overline{\omega}_2}{\omega_{ln}} \\
   &\omega_{ln} \equiv \exp(\frac{1}{\lambda} \int_0^\infty \frac{\alpha^2 f(\omega)}{\omega} \log(\omega) d\omega) \\
   &\overline{\omega}_2 \equiv \frac{\int_0^\infty \alpha^2 f(\omega) \omega d\omega}{\int_0^\infty \alpha^2 f(\omega)/\omega d\omega}
 \end{align}  

  To study large amounts of stoichiometries in our model structures and try to maximize T$_c$, we take advantage of the established correlation between certain hydride electronic properties and the isotropic Eliashberg T$_c$ \cite{belli2021strong}, which yields a method to roughly estimate T$_c$. The T$_c$ estimator takes the form
 \begin{align}
   \label{eq:tc}
   & T^{net}_c \approx 750 \Phi - 85 \quad (\text{Kelvin}) \\
   & \Phi \equiv \phi^{net}_{iso} \times H_f \times \sqrt[3]{\text{DOS}_{H,rel}(E_F)}\\
   & H_f \equiv \frac{\text{\# of H forming network}}{\text{\# of atoms in unit cell}}\label{eq:Hf}\\
   & \text{DOS}_{H,rel}(E_F)  \equiv \frac{\text{DOS}_H(E_F)}{\text{DOS}_{tot}(E_F)} \\
 \end{align} where $\phi^{net}_{iso}$ is the critical hydrogen networking isovalue \cite{belli2021strong} which roughly measures the bonding strength/connectivity of the hydrogen sublattice. It is determined by the isovalue in which the electron localization function (ELF) of electrons centered on different hydrogen atoms begin to overlap. H$_f$ is the fraction of hydrogen in the unit cell forming a network with overlapping ELF from different hydrogen, and DOS$_{H,rel}(E_F)$ is the relative contribution of hydrogen states to the density of states at E$_F$. T$_c^{net}$ has been established to reproduce the computed isotropic Eliashberg T$_c$ of hundreds of hydrides within $\pm 60$ K \cite{belli2021strong}, which we find acceptable to identify material candidates to look further into.

\paragraph*{Acknowledgements. } This research was supported by the U.S. National Science Foundation (NSF, DMR-2104881; R.H.), the U.S. Department of Energy-National Nuclear Security Administration (DOE-NNSA) through the Chicago/DOE Alliance Center (DE-NA0003975; A.D., R.H.), the DOE Office of Science (DE-SC0020340, R.H.), and NSF SI2-SSE (Grant 1740112; H.P.).



\bibliography{quaternaries_main_v9_submit.bbl}

\end{document}